\documentclass[lettersize,journal]{IEEEtran}
\usepackage{amsmath,amsfonts}
\usepackage{algorithmic}
\usepackage{algorithm}
\usepackage{array}
\usepackage[caption=false,font=normalsize,labelfont=sf,textfont=sf]{subfig}
\usepackage{textcomp}
\usepackage{stfloats}
\usepackage{url}
\usepackage{verbatim}
\usepackage{graphicx}
\usepackage{multirow}%
\usepackage{xcolor}%
\usepackage{textcomp}%
\usepackage{manyfoot}%
\usepackage{booktabs}%
\usepackage{listings}%
\usepackage[backend=bibtex,bibstyle=ieee,citestyle=numeric-comp]{biblatex}
\bibliography{Bibliography}
\begin{document}

\title{Unpaired Test for the Comparison of Frequency Response Functions Groups}

\author{Vittorio Lippi%~\IEEEmembership{Staff,~IEEE,}
        % <-this % stops a space
\thanks{Vittorio Lippi is with Institute of Digitalization in Medicine, Faculty of Medicine and Medical Center, University of Freiburg, 79106 Freiburg, Germany; Clinic of Neurology \textit{and}, Neurophysiology, Medical Centre, University of Freiburg, Faculty of Medicine, University of Freiburg, Breisacher Straße 64, 79106 Freiburg im Breisgau, Germany}
\thanks{Email: vittorio.lippi@uniklinik-freiburg.de}
\thanks{This paper is additional material to \cite{Lippi2024forthcoming} }}% <-this % stops a space
%\thanks{Manuscript received April 19, 2021; revised August 16, 2021.}}

%\markboth{Additional Material}%


\maketitle

\begin{abstract}
The frequency response function (FRF) is a typical way to describe the outcome of experiments where posture control is perturbed with an external stimulus. The FRF is an empirical transfer function between an input stimulus and the induced body segment sway profile, represented as a vector of complex values associated with a vector of frequencies. This work proposes an unpaired test based on bootstrap to compare the averages the outcome of posture control experiments. 
\end{abstract}

\begin{IEEEkeywords}
Frequency response function, Bootstrap, unpaired.
\end{IEEEkeywords}

\section{Introduction}
\textbf{The frequency response function (FRF)} is a usual way to describe the outcome of experiments in posture control literature. In particular, the FRF is an empirical transfer function between an input stimulus and the induced body movement with an additional step of averaging the value at some sets of frequencies, as explained in detail in section \ref{FRF}. By definition, the FRF is a complex function of frequency. When statistical analysis is performed to assess differences between groups of FRFs (e.g., obtained under different conditions or from a group of patients and a control group), the FRF's structure should be considered. Usually, the statistics are performed by defining a scalar variable to be studied, such as the norm of the difference between FRFs, or considering the components independently that can be applied to real and complex components separately\cite{lippi2023human,Akcay2021}, in some cases both approaches are integrated, e.g., the comparison frequency-by-frequency is used as a post-hoc test when the null hypothesis is rejected on the scalar value\cite{lippi2020body}. The two components of the complex values can be tested with multivariate methods such as Hotelling's T2 as done in \cite{asslander2014sensory} on the averages of the FRF over all the frequencies, where a further post hoc test is performed applying bootstrap on magnitude and phase separately. The problem with the definition of a scalar variable as the norm of the differences or the difference of the averages in the previous examples is that it introduces an arbitrary metric that, although reasonable, has no substantial connection with the experiment unless the scalar value is assumed a priori as the object of the study as in \cite{lippi2020human,robovis21} where a human-likeness score for humanoid robots is defined on the basis of FRFs difference with the aim to benchmark humanoid robots \cite{torricelli2020benchmarking,Lippi2019}. On the other hand, testing frequencies (and components) separately does not consider that the FRF's values are not independent and applying corrections for multiple comparisons, e.g., Bonferroni, can result in a too-conservative approach destroying the power of the experiment. In order to properly consider the nature of the FRF, a method oriented to complex functions should be used. In \cite{Lippi2023} a preliminary method based on random field theory inspired by \cite{pataky2016region} was presented: to take into account the two components (imaginary and real) as two independent variables, the fact that the same subject repeated the test in the two conditions, a 1-D implementation of the Hotelling T2 is used as presented in\cite{pataky2014vector} but applied in the frequency domain instead of the time domain.

 In the specific case of FRFs from posture control experiments, the sample itself, by definition, provides a predefined set of frequencies (as it will be clear in \S \ref{FRF}). This suggests that all the FRFs in the samples can be conveniently transformed to time domain signals that consistently represent all the components of the FRFs. On this basis, the present work proposes a bootstrap method for testing the difference detween the means of two groups of FRFs. The method is implemented by transforming the FRF in the time domain, obtaining a \textit{pseudo-impulse-response} (PIR) function on which such bands can be defined and computed. The results of the tests in the frequency domain are then presented and discussed.

\textbf{This paper} Describes an extension to the FRFs library \cite{LippiFRF24} consisting in a new function to perform unpaired tests. The library was originally developed to define confidence and predictions band as extensively described in \cite{Lippiforthcoming}.  In the \textit{Methods} section, a definition of the FRF used in posture control literature is provided for reference, and then the Bootstrap method is described. then an example is presented with data from the paper \cite{Lippi2024forthcoming}. The Results section presents the results of the tests. In the last section, \textit{The Code}, there are links to the repository to download the Matlab \cite{MATLAB:2019b}source and instructions to run the functions computing the bands and running tests.

\section{Methods}
\subsection{The FRF}
\label{FRF}
\begin{figure*}
\centering
\includegraphics[width=1.00\textwidth]{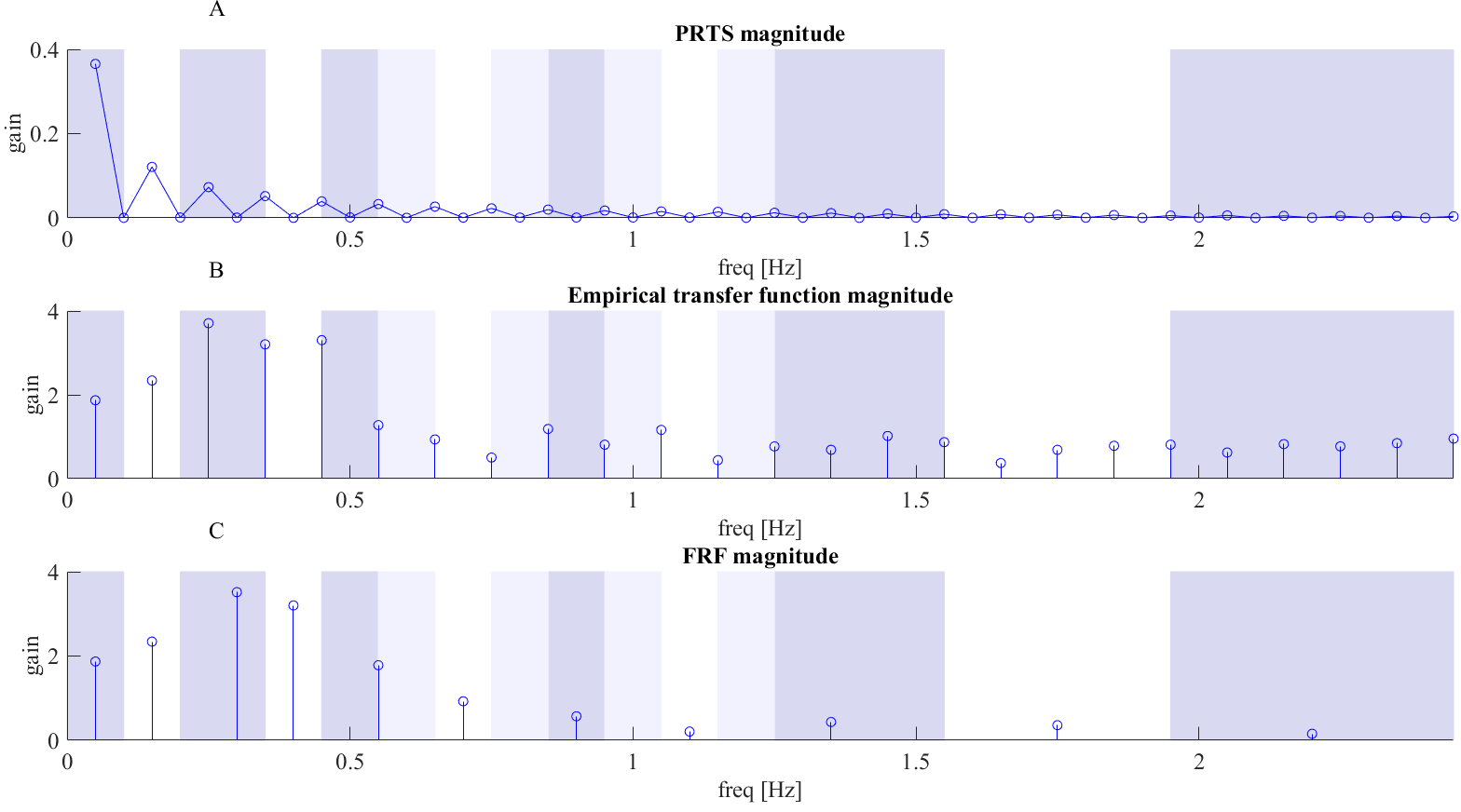}
\caption{An example of FRF. On top of the magnitude of the PRTS DFT, notice the \textit{comb} structure with peaks alternated to zeroes. In the middle, the magnitude of a sample empirical transfer function is defined on the frequencies associated with non-zero values in the input PRTS. Below, the FRF is obtained by averaging the transfer function over different ranges of frequencies. The ranges are represented in the three plots with bands of different colors, i.e., white and blue, and, as the ranges are overlapping, the light blue bands belong to two contiguous ranges. In the final FRF, the average over a range of frequencies is associated with the mean of the frequencies in such range. Although the plots show just the magnitude for ease of reading, the average is computed in a complex domain, and the FRF is a complex function.}
\label{fig:FRF}
\end{figure*}

The frequency response function, FRF, is an empirically computed transfer function defined between an input and output. In posture control experiments, the input is a stimulus, such as a support surface tilt with a specific profile, and the output is the body sway or the sway of a body segment. In this work, the FRF is defined as proposed in \cite{peterka2002sensorimotor}, in detail:
\begin{enumerate}
\item The input has a pseudo-random ternary signal profile (PRTS) composed of segments at constant speed $(0,+s,-s)$ where $s$ is a fixed speed value usually set to get a specific amplitude of the position profile (e.g., $1^\circ$ peak-to-peak tilt).
\item The PRTS has a power spectrum characterized by peaks separated by zones with zero power, as shown in Fig. \ref{fig:FRF}. The response is computed on such frequencies: Sway responses are averaged across all PRTS sequence repetitions across subjects, discarding the first cycle of each trial to avoid transient response effects. Spectra of the stimulus and body sway in space are computed using the Fourier transform. Finally, frequency response functions are computed as cross-power spectrum $G_{xy}(f)$ divided by the stimulus power spectra $G_{yy}(f)$ that is $G_{xy}/G_{yy}$.
\item in the examples presented in this paper, the signals are sampled at 50 Hz.
\item The values of the obtained transfer function are averaged over bands of frequencies as shown in Fig. \ref{fig:FRF}, with the resulting FRF being represented by a vector of 11 complex values\footnote{in the original formulation \cite{peterka2002sensorimotor} the frequencies were 22 corresponding to a stimulus that was twice long, i.e. same profile but $100\%$ slower} associated to the frequencies $\varphi=[ 0.05,\: 0.15,\: 0.3,\: 0.4, \:0.55,\: 0.7,\: 0.9,\: 1.1, \:1.35, \:1.75, \:2.2 ]$.
\end{enumerate}
in summary, such a definition of FRF differs slightly from a usual discrete transfer function because it is defined on the \textit{custom} frequencies $\varphi$.

\subsection{The Bootstrap Method}
\begin{figure*}[tb!]
\centering
\includegraphics[width=1.00\textwidth]{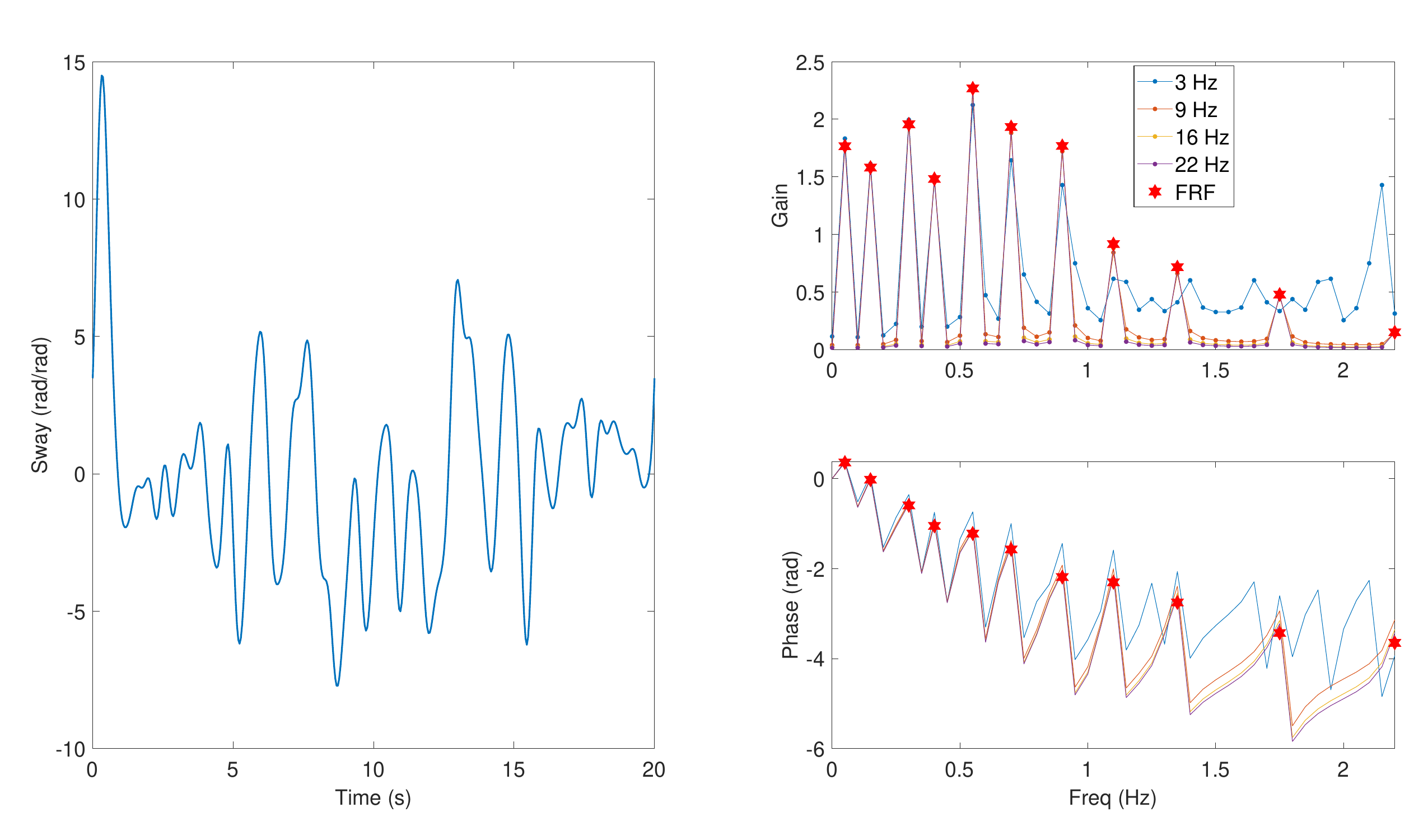}
\caption{The PIR on the left, the FRF (red stars), and the DFT of the PIR (colored lines). The PIR is computed with eq. \ref{pseudo}. Different sample times are tested to reconstruct the FRF through a DFT. The period of the PIR is defined as the inverse of the greatest common divisor of the frequencies in $\varphi$; the sample time used in the examples is set to ten times the highest frequency in $\varphi$, i.e., 22 Hz. Notice how the gain tends to converge to zero between the peaks}
\label{Pseudopulse}
\end{figure*}

The test is performed comparing two groups of FRFs. A FRF is denoted as $H_{f}$. As the FRF is a \textit{sort of} transfer function, it can be considered the Fourier transform of a real signal representing the system's impulse response. It will be referred to as \textit{pseudo-impulse-response}, PIR, can be computed as
\begin{equation}
\label{pseudo}
x(t)= \sum^{M}_{k=1} \Re(H_{k}) \cos(2\pi F_{k} t) + \Im(H_{k})\sin(2\pi F_{k} t)
\end{equation} %M=11
In the present work, $M=11$ following the structure of the FRF presented in \cite{lippi2023human,lippi2020human}, in the original formulation \cite{peterka2002sensorimotor}, the FRF had 22 components. The frequency $F_k$ is $k$th component of the vector $\varphi$. The PIR $x(t)$ can be seen as a vector of coefficients defining the FRF as the time $t$ defined on a discrete range. The period of the PIR is defined as the inverse of the greatest common divisor of the frequencies in $\varphi$. Figure \ref{Pseudopulse} shows how the FRF can be reconstructed by applying the DFT to the PIR with different numbers $T$ of time samples. The sample time in the examples is chosen to have ten times the highest frequency in $\varphi$. The PIR is equivalent to the FRF as it can be transformed back to the frequency domain by a DFT, leading to a complex function that is equal to the FRF on the frequencies in $\varphi$ and zero elsewhere as shown in Fig.\ref{Pseudopulse}.
The statistical analysis of the PIRs allows an analysis of the FRF distribution that combines all the frequencies together.
Given the set of N PIRs, the mean and the standard deviation can be estimated at each sample time to describe the distribution as follows:
\begin{align}
\label{variability}
\bar{p}(t)= \frac{1}{N} \sum^{i=1}_{N} p_i(t) \\
\hat{\sigma}_p(t) = \sqrt{\frac{1}{N-1}\sum^{i=1}_{N} \left|p_i(t)-\bar{p}(t) \right|^2}\\
\end{align}

\textbf{The confidence band on the difference between the mean of the groups} can then be defined to obtain given the desired confidence level. The function $p(t)$ considered is the difference between the averages of the two groups. This means that the $N$ samples in the equations above are produced with bootstrap repetitions. Given the desired confidence level $\alpha\%$ the constant $C_p$ is defined to obtain the probability:
\begin{equation}
P\left[ \max\limits_{t} \left( \frac{|p(t) - \hat{p}(t)|}{\hat{\sigma}_{p}(t)} \right) \leq C_c\right] = \frac{\alpha}{100}
\label{confidenceeq}
\end{equation}
The $\alpha\%$ confidence band for $\hat{x}(t)$ is then
\begin{equation}
\label{confidenceconstant}
\hat{x}(t) \pm C_c \cdot \hat{\sigma}_{p}(t)
\end{equation}
\textbf{The bootstrap} is used to determine $C_c$ . Approximated versions of the probabilities in eq.\ref{confidenceeq} are obtained using empirical distributions produced by resampling the sample set. The constants $C_c$ is set so that the approximated probability is as close as possible to the desired confidence $\alpha\%$. Specifically eq. \ref{confidenceeq} has the following bootstrap approximation:
\begin{equation}
\frac{1}{B}\sum\limits_{b=1}^{B}\left[I \left( \max\limits_{t} \left( \frac{|\hat{p}^b(t) - \bar{p}(t)|}{\hat{\sigma}^{b_{N}}_{p}(t)} \right)  \leq C_c\right)\right]
\label{confidencebootstrap}
\end{equation}
where I(E) is 1 or 0 according to condition E is or is not verified, respectively. Eq. \ref{confidencebootstrap} is the average, over the $B$ bootstrap replications, of the proportion of the original data curves whose maximum standardized deviation from the bootstrap mean is less than or equal to $C_p$. The superscript $^b$ addressed that the quantity is computed on the basis of the resampled set and $^{b_{N}}$ means that the quantity is computed with a nested bootstrap loop. In fact $\hat{\sigma}^{b_{N}}_{p}(t)$ is based on the resample set at every iteration of the bootstrap as recommended in \cite{hall1991two}. Such ``pivotization'' also allows for null hypothesis testing without having to simulate the distribution produced by the null hypothesis \cite{davison2003recent}. This is implemented with a nested bootstrap loop. The number $B$ is set to be relatively large with a trade-off between the accuracy and the computational time. Different indications about the required $B$ are discussed in \cite{zoubir1998bootstrap}. . In previous posture control experiments \cite{lippi2023human} and \cite{lippi2020body}, the components were tested independently, and several resampled sets in the order of $B=10^4$ produced acceptable results with groups of $7$ and $36$ subjects, respectively. The specific bootstrap included a nested resampling $B_{NEST}=200$ used to estimate the variance. 

\subsection{Sample Sets and Tests}
\label{section:tests}
The data are the ones from the paper \cite{Lippi2024forthcoming}. In particular the COM sway response is shown as an example (in the paper also head sway is tested). Given the same condition (eyes-open or eyes-closed) the groups of IPS and PSP patients are compared with the control group.

\subsection{Visualizing the Results in Frequency Domain}
Using the bands defined on the PIRs for tests provides a result visualized in the time domain without immediate interpretations. Mapping the result in the frequency domain allows for considerations about how the different distribution sets (subjects groups or conditions) differ in specific frequency ranges (e.g., in \cite{joseph2014contribution} where bands like \textit{low frequencies}, \textit{middle frequencies} and \textsl{high frequencies} are defined manually). A residual can be defined as the difference between a PIR $x_i(t)$that exceeds the bands:
\begin{equation}
\scriptstyle
r(t)=\left\{\begin{array}{cl} 
\scriptstyle  x_i(t) -(\hat{x}(t) + C_c \cdot \hat{\sigma}_{x}(t)),& \scriptstyle  x_i(t) \geq \hat{x}(t) + C_c \cdot \hat{\sigma}_{x}(t)\\
\scriptstyle  0,& \scriptstyle |\hat{x}(t) - x_i(t)| \leq C_c \cdot \hat{\sigma}_{x}(t)\\
\scriptstyle x_i(t) -(\hat{x}(t) - C_c \cdot \hat{\sigma}_{x}(t)),& \scriptstyle x_i(t) \leq \hat{x}(t) - C_c \cdot \hat{\sigma}_{x}(t)\end{array}\right.
\label{eq:residual}
\end{equation}
The DFT of $r(t)$, specifically on the frequencies $\varphi$, provides a frequency domain plot of how the sample exceeded the bands. While the complex function does not have an easy interpretation, its gain or power spectrum can be easily used to make the above considerations. The DFT of $r(t)$ is shown in Figures  \ref{fig:COM2_IPS_PSP4}, \ref{fig:COM2_IPS_PSP7}, \ref{fig:COM2_IPS_PSP8}, and \ref{fig:COM2_IPS_PSP16} (C).
\section{Results}
\label{section:results}
\begin{figure*}
	\centering
		\includegraphics[width=1.00\textwidth]{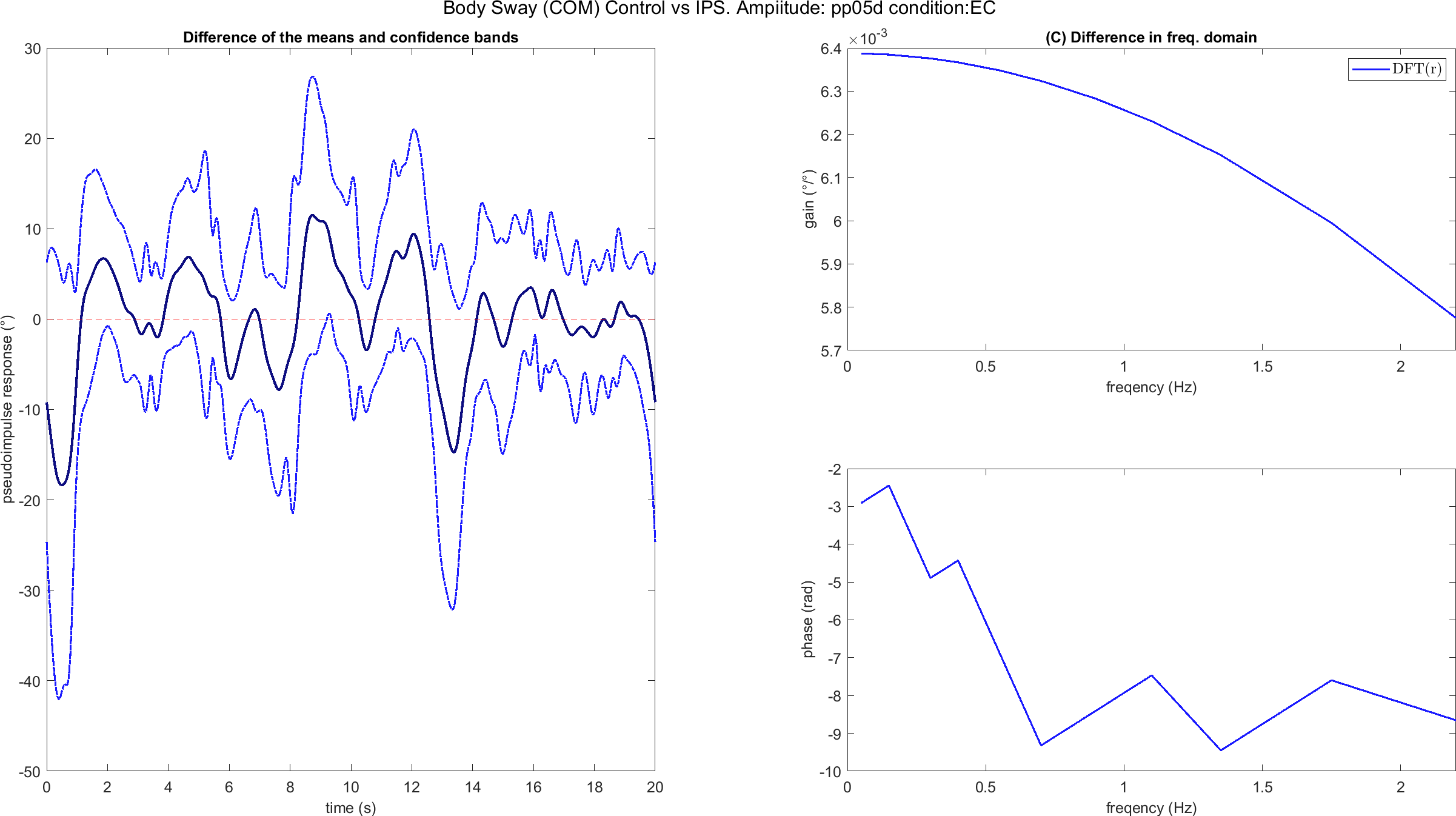}
	\label{fig:COM2_IPS_PSP4}
		\caption{Control Group vs IPS, amplitude 0.5 degrees, condition eyes closed. LEFT: The $95\%$ confidence bands on the difference between the average of the two groups is compared with the abscissa axis, i.e. $x(t)=0$ that represents the null hypothesis that there is no difference between the averages. As the axis exceeds the confidence bands at around $t=1s$ the null hypothesis is rejected. RIGHT: the Fourier transform of the residuals reveals a difference in the average responses of the two groups mainly expressed at low frequency.}
\end{figure*}

\begin{figure*}
	\centering
		\includegraphics[width=1.00\textwidth]{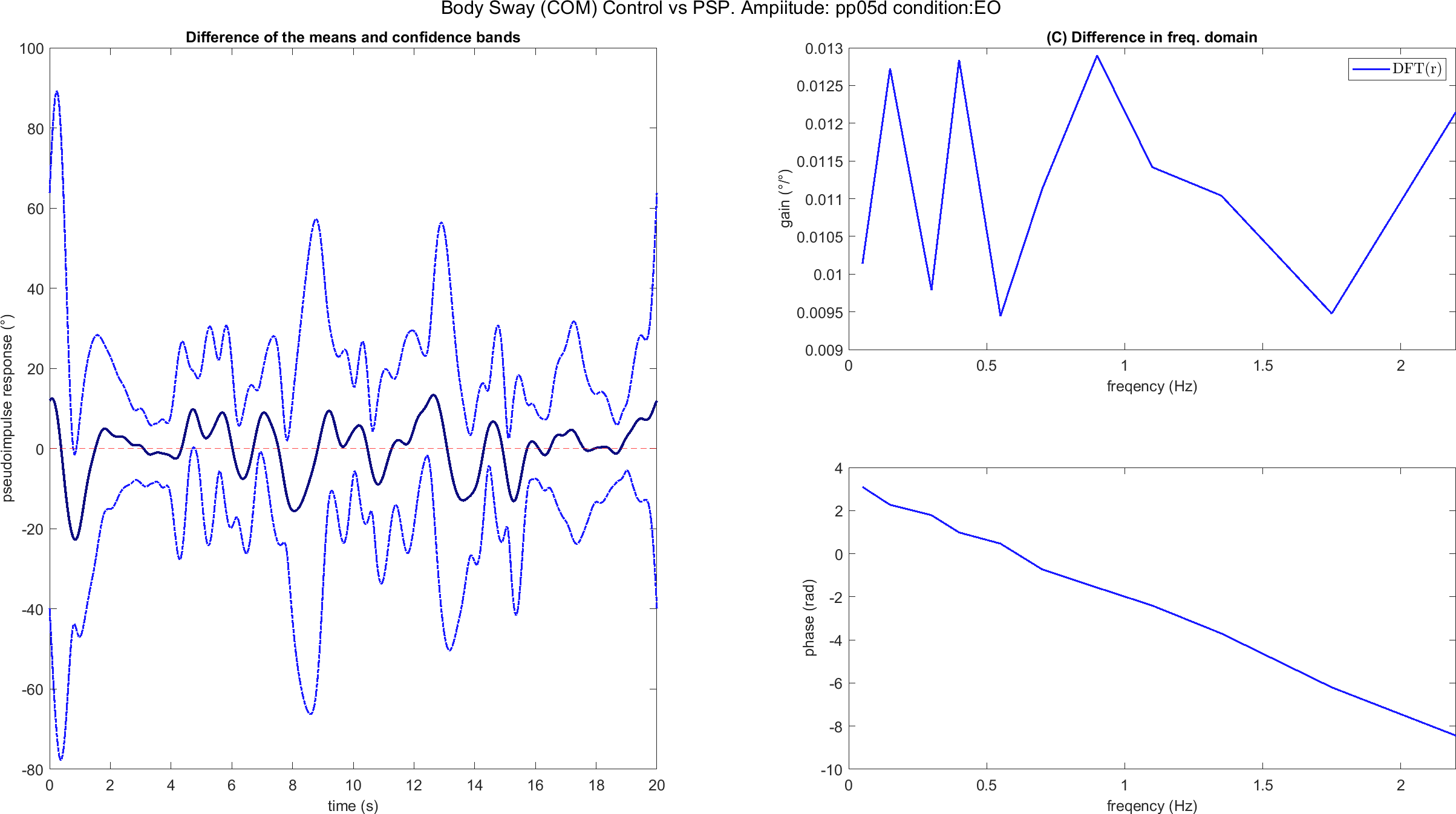}
		\caption{Control Group vs PSP, amplitude 0.5 degrees, condition eyes open. LEFT: The $95\%$ confidence bands on the difference between the average of the two groups is compared with the abscissa axis, i.e. $x(t)=0$ that represents the null hypothesis that there is no difference between the averages. As the axis exceeds the confidence bands at around $t=1s$ and at $t=5s$  the null hypothesis is rejected. RIGHT: the Fourier transform of the residuals reveals a difference in the average responses of the two groups expressed over the whole spectrum of frequencies.}
	\label{fig:COM2_IPS_PSP7}
\end{figure*}

\begin{figure*}
	\centering
		\includegraphics[width=1.00\textwidth]{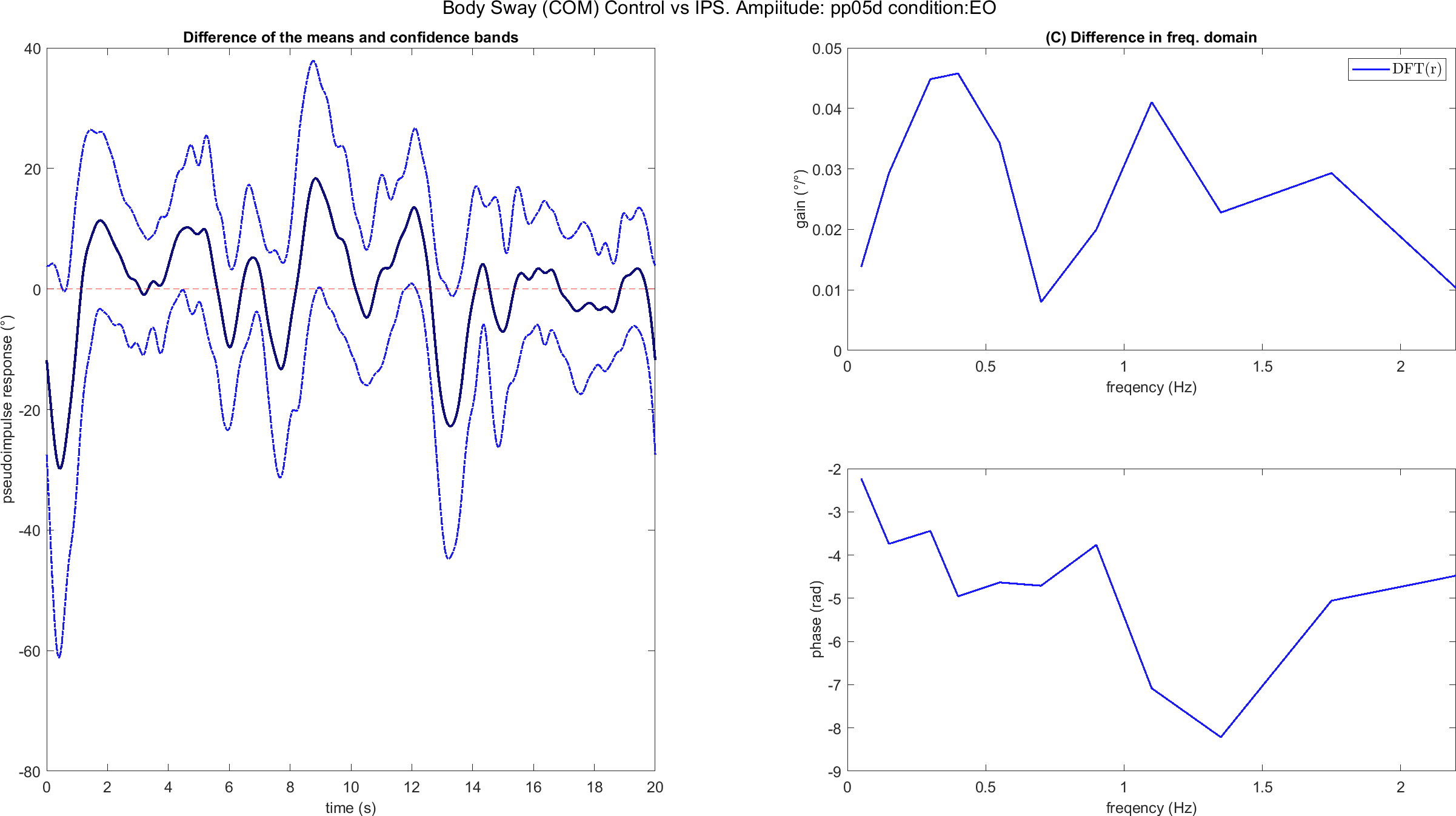}
	\caption{Control Group vs IPS, amplitude 0.5 degrees, condition eyes open. LEFT: The $95\%$ confidence bands on the difference between the average of the two groups is compared with the abscissa axis, i.e. $x(t)=0$ that represents the null hypothesis that there is no difference between the averages. As the axis exceeds the confidence bands at around $t=1s$, $t=5s$,  $t=9s$,   $t=12s$  and at $t=14s$  the null hypothesis is rejected. RIGHT: the Fourier transform of the residuals reveals a difference in the average responses of the two groups expressed over the whole spectrum of frequencies.}
	\label{fig:COM2_IPS_PSP8}
\end{figure*}

\begin{figure*}
	\centering
		\includegraphics[width=1.00\textwidth]{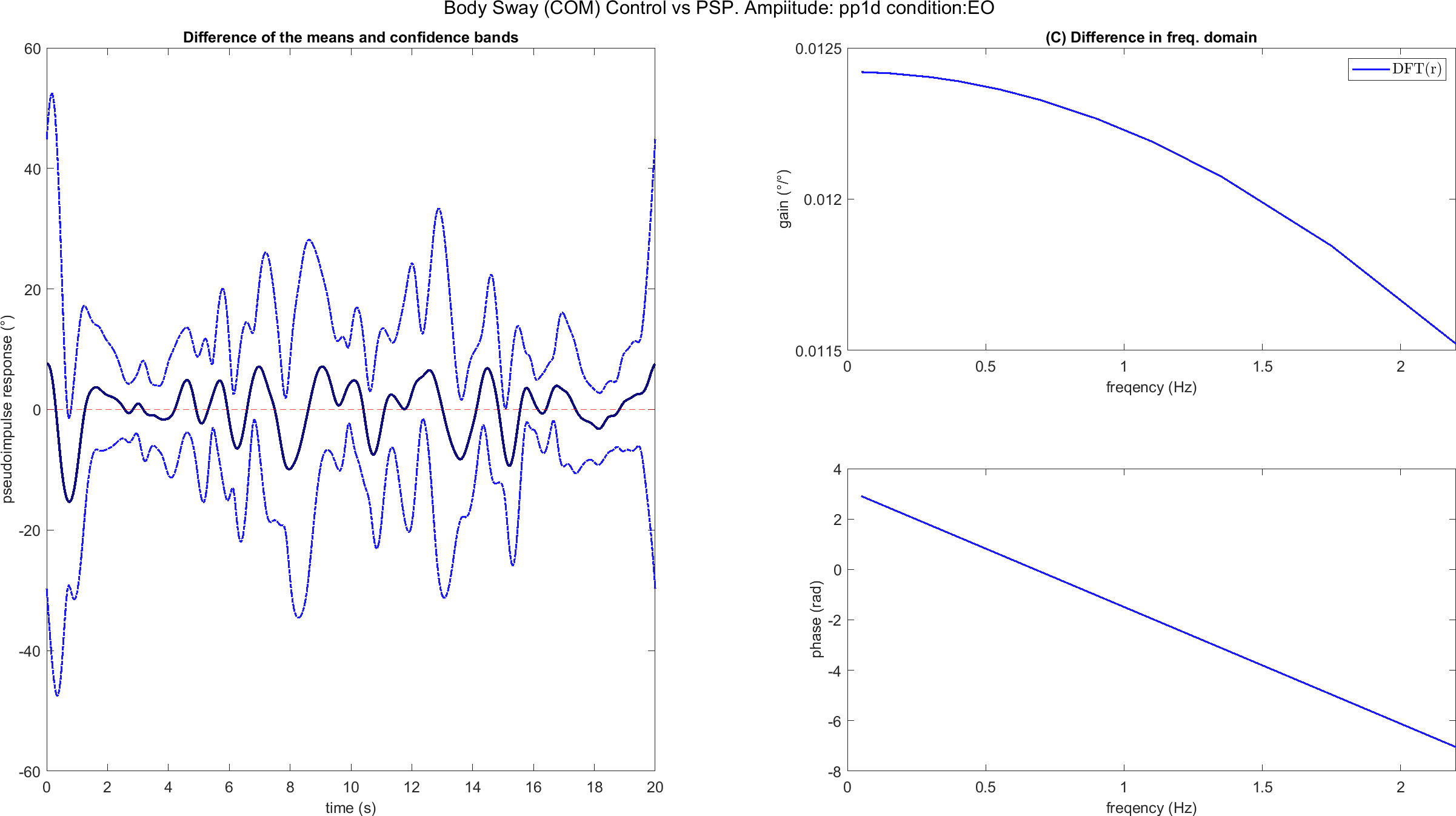}
	\caption{Control Group vs PSP, amplitude 1 degre, condition eyes open. LEFT: The $95\%$ confidence bands on the difference between the average of the two groups is compared with the abscissa axis, i.e. $x(t)=0$ that represents the null hypothesis that there is no difference between the averages. As the axis exceeds the confidence bands at around $t=1s$ and at $t=15s$.  The null hypothesis is rejected. RIGHT: the Fourier transform of the residuals reveals a difference in the average responses of the two groups expressed mostly over low pass frequencies.}
	\label{fig:COM2_IPS_PSP16}
\end{figure*}

The sample tests described in \S \ref{section:tests} show the application of the method. The test is performed comaring the confidence bands constructed with the bootstrap method with the x-axis, that represents the null hypothesis, i.e. $p(t)=0$. if the axis is not included in the confidence bands the null hypothesis is rejected as shown in the Figures  \ref{fig:COM2_IPS_PSP4}, \ref{fig:COM2_IPS_PSP7}, \ref{fig:COM2_IPS_PSP8}, and \ref{fig:COM2_IPS_PSP16} (Left).

\section{Discussion and future work}
The proposed example showed how to implement an unpaired test to check the difference between two group of subjects. The tests performed in \cite{Lippi2024forthcoming} are performed with the described method. Applying the bootstrap procedure on the PIR function, that is real and defined in time domain, allowed for a straightforward application of 1-D statistics. Converting back the residuals to frequency domain allows a comparison between the original FRFs and the result of the tests (e.g. localizing the frequencies where the average differs most from the null hypothesis. The versatility of the bootstrap easily allows for extensions, for example in forthcoming work the probability that a FRF belongs to a certain set is approximated with an extension of the presented method \cite{Lippi2024a,newFRFpaper}.
\onecolumn
\section{The Code}
\vspace{-60ex}
\textbf{The unpaired test} is performed with the following function
\vspace{-20ex}
\begin{verbatim}
function [avg,sigma,band,Cc,chist,values] = FRF_ConfidenceBandDifference(FRF1,FRF2,phi,sample_time,alpha,B,Bs)
%[AVG,SIGMA,band,CC,CHIST,VALUES] =
%FRF_CONFIDENCEBANDDIFFERENCE(FRFS1,FRFS2,PHI,SAMPLE_TIME,B)
% Confidence bands on the difference between the means of two groups FRF1
% andd FRF2.
%
% B is the number of bootstrap repetitions
% Bs is the number of bootstrap repetitions used to estimate STD
%
% avg is the difference between average PIRs of the groups, sigma is the
% STD of the stat, band a two row matrix with the boundaries of
% the band. Cp is the threshold constant obtained by the bootstrap.  FRFS
% is a matrix where each row represents a FRF of the set, phi is the vector
% of frequencies  and SAMPLE_TIME is the sample time of the PIRs. Chist is
% a vector representing the cumulative histogram for the values  returned
% in VALUES.

sf=1/sample_time;

N1=size(FRF1,1); %number of FRFs

x1=FRF_pseudoimpulse(FRF1(1,:),phi,sf);
ns=length(x1);
y1=zeros(N1,ns);
y1(1,:)=x1;

for i=2:N1
    [x,t]=FRF_pseudoimpulse(FRF1(i,:),phi,sf);
    y1(i,:)=x;
end

N2=size(FRF2,1); %number of FRFs

x1=FRF_pseudoimpulse(FRF2(1,:),phi,sf);
ns=length(x1);
y2=zeros(N1,ns);
y2(1,:)=x1;

for i=2:N2
    [x,t]=FRF_pseudoimpulse(FRF2(i,:),phi,sf);
    y2(i,:)=x;
end



xm=mean(y1)-mean(y2);

sSTAT=zeros(Bs,ns);
for b=1:Bs
    resamp1=randi(N1,1,N1);
    resamp2=randi(N2,1,N2);
    yb1=y1(resamp1,:);
    yb2=y2(resamp2,:);
    
    sSTAT(b,:)=mean(yb1)-mean(yb2);
end
sx=std(sSTAT);


STAT=zeros(1,B);


for b=1:B %GENERATE THE HISTOGRAM
    resamp1=randi(N1,1,N1);
    resamp2=randi(N2,1,N2);
    yb1=y1(resamp1,:);
    yb2=y2(resamp2,:);
    
    xb=mean(yb1)-mean(yb2);
    
    for b2=1:Bs
        resampb1=randi(N1,1,N1);
        resampb2=randi(N2,1,N2);
        ybb1=yb1(resampb1,:);
        ybb2=yb2(resampb2,:);
        
        sSTAT(b2,:)=mean(ybb1)-mean(ybb2);
    end
    
    sb=std(sSTAT);
    db=max(abs(xm-xb)./sb);
    STAT(b)=db;
end


%% Histogram

STAT=sort(STAT);
%H=histogram(STAT,1000,'Normalization','cdf','Edgecolor','none')
[chist, values] = histcounts(STAT,1000,'Normalization','cdf');
Cc=values(find(chist>alpha,1,'first'));

avg=xm;
sigma=sx;

band=[avg+Cc*sigma;avg-Cc*sigma];

end
\end{verbatim}

where \texttt{avg} is the average of the differences of the mean PIRs of the two groups over bootstrab repetitions, \texttt{sigma} is the measure of variation $\hat{\sigma}_x(t)$, \texttt{band} a two row matrix with the boundaries of the bands $C_c\cdot \hat{\sigma}_x(t)$. \texttt{Cc} is the constant $C_c$ obtained by the bootstrap. The constant $C_c$ is computed by ordering the values of the statistics (maximum difference between the means normalized by the standard deviation, \texttt{STAT}) produced by each bootstrap repetition and computing a cumulative histogram on which the constant is chosen to obtain the desired confidence $\alpha$. \texttt{FRF1} and \texttt{FRF2} are matrices where each row represents a FRF of the set, \texttt{phi} is the vector of frequencies $\varphi$ and \texttt{sample\_time} is the sample time of the PIRs, set by default to ten times the highest frequency in $\varphi$. \texttt{chist} is a vector representing the cumulative histogram for the values returned in \texttt{values}.

\textbf{Computing the PIR} is performed by the service function FRF\_pseudoimpulse.
\begin{verbatim}
function [x,t]=FRF_pseudoimpulse(y,F,sf)
% manually computes the inverse of the FRF chosing the frequencies (11
% components, maurer's version)
bf=double(gcd(sym(F))); %this defines the period


step=(1/sf);
t=0:step:(1/bf);
n=length(F);
x=0;
for i=1:n
    x=x+real(y(i))*cos(2*pi*F(i)*t)-imag(y(i))*sin(2*pi*F(i)*t);
end
\end{verbatim}
\texttt{FRF} is a single FRF, \texttt{Freq} is the vector $\varphi$ of frequencies and \texttt{sf} is the sample frequency. The function returns the PIR \texttt{x} and the time vector \texttt{t}.

\textbf{The source code is available} in the repository: https://github.com/mcuf-idim/FRF-statistics

\twocolumn

\printbibliography
\end{document}